\documentstyle[twoside,fleqn,espcrc2,epsfig]{article}

\def\spose#1{\hbox to 0pt{#1\hss}}
\def\ltapprox{\mathrel{\spose{\lower 3pt\hbox{$\mathchar"218$}}
 \raise 2.0pt\hbox{$\mathchar"13C$}}}
\def\gtapprox{\mathrel{\spose{\lower 3pt\hbox{$\mathchar"218$}}
 \raise 2.0pt\hbox{$\mathchar"13E$}}}
\def\inapprox{\mathrel{\spose{\lower 3pt\hbox{$\mathchar"218$}}
 \raise 2.0pt\hbox{$\mathchar"232$}}}

\newcommand{\sea}{\mbox{\scriptsize sea}}

\newcommand{\val}{\mbox{\scriptsize val}}

\newcommand{\SW}{\mbox{\scriptsize SW}}

\newcommand{\etal}{{\em et al.}}

\newcommand{\NPB}{Nucl.Phys.B}
\newcommand{\PLB}{Phys.Lett.B}
\newcommand{\PRD}{Phys.Rev.D}

\newcommand{\be}{\begin{equation}}
\newcommand{\ee}{\end{equation}}
\newcommand{\bea}{\begin{eqnarray}}
\newcommand{\eea}{\end{eqnarray}}

\newcommand{\csw}{c_{\rm sw}}

\newcommand{\ksea}{\mbox{$\kappa^{\rm sea}$}}
\newcommand{\kval}{\mbox{$\kappa^{\rm val}$}}

\newcommand{\plus}{\makebox[15pt][c]{$+$}}
\newcommand{\minus}{\makebox[15pt][c]{$-$}}
\newcommand{\err}[2]{\raisebox{0.08em}{\scriptsize
                          {$\;\begin{array}{@{}l@{}}
                          \plus\makebox[1.0em][r]{#1} \\[-0.12em]
                          \minus\makebox[1.0em][r]{#2}
                        \end{array}$}}}

\title{Light hadron spectroscopy with $O(a)$ improved dynamical fermions.}

\author{
M.~Talevi\address{
Department of Physics \&\ Astronomy, University of Edinburgh,
The King's Buildings, EH9 3JZ (UK)}
\thanks{Talk presented at Lattice '98, XVI International Symposium on Lattice Field Theory, Boulder, CO (USA), July 1998. 
Supported by PPARC through grant GR/L22744}, 
{\em for the UKQCD Collaboration}
}

\begin{document}

\begin{abstract}
We present results for the hadron spectrum and static quark potential
from a simulation with two flavours of $O(a)$ improved dynamical
Wilson fermions at $\beta=5.2$.  We address the issues of sea quark
dependence of observables and finite-size effects.
\end{abstract}

\maketitle

Quenching is the only non-systematically improvable approximation in lattice
QCD.  Moreover, sea-quark effects are to be taken into account if we
are to make theoretical predictions on such phenomena as $\eta'-\pi$ splitting
and string breaking.  Dynamical simulations are computationally
very expensive and control of systematic effects is thus of the utmost 
importance.  The improvement counterterm $c_{\SW}(g_0^2)$ for the SW-improved 
action has been computed non-perturbatively, thus yielding full 
$O(a)$ improvement with $N_f=2$ dynamical fermions \cite{Jansen}.  

The simulation parameters are summarized in tab.~\ref{tab:params}.  
For each volume and each value of $\kappa_{\sea}$ we have simulated
valence quarks with an appropriate set of $\kappa$'s to exploit  
the full $(\kappa_{\sea},\kappa_{\val})$ plane for 
``strange'' physics \cite{UKQCD_C176}. At $\beta=5.2$ the PS/V mass ratios
for the relevant $\kappa$'s are shown in tab.~\ref{tab:masses}.
The value of $c_{\SW}$ used differs from the final estimate of 
the {\sc Alpha} Collaboration \cite{Jansen}, but since a notable
part of the simulation had already been carried out with this improvement
coefficient, UKQCD have decided to continue the simulation,
concentrating on finite-size effects and looking for signs of sea-quark effects
in both the static quark potential and the light hadron spectrum.  
The complete removal of $O(a)$ artefacts is thus not the highest priority 
in this work.
Details of the simulation and results can be found in \cite{UKQCD_C176}, while
we refer the reader to \cite{Kenway} for an up-to-date review on
spectroscopy.

\begin{table}[t]
\begin{tabular}{rrrrr}
\hline
$\beta$ & $c_{\SW}$ & $L^3\cdot T$ & $\kappa_{\sea}$  & Conf \\ \hline \hline
 5.2 &  1.76 & $8^3\cdot 24$  &  $0.1370$  & 78 \\
     &       &                &  $0.1380$  & 100 \\
     &       &                &  $0.1390$  & 100 \\
     &       &                &  $0.1395$  & 60 \\ \hline 
 5.2 &  1.76 & $12^3\cdot 24$ &  $0.1370$  & 151 \\
     &       &                &  $0.1380$  & 151 \\
     &       &                &  $0.1390$  & 151 \\
     &       &                &  $0.1395$  & 121 \\
     &       &                &  $0.1398$  & 98 \\ \hline 
 5.2 &  1.76 & $16^3\cdot 24$ &  $0.1390$  & 90 \\
     &       &                &  $0.1395$  & 100 \\
     &       &                &  $0.1398$  & 69 \\ \hline 
\end{tabular}
\caption{Simulation parameters.}
\vspace{-1.0truecm}
\label{tab:params}
\end{table}

\section{Static quark potential}

\begin{figure}[t]
\hspace{-2.0truecm}
\vspace{-4.0truecm}
\begin{center}
\epsfig{figure=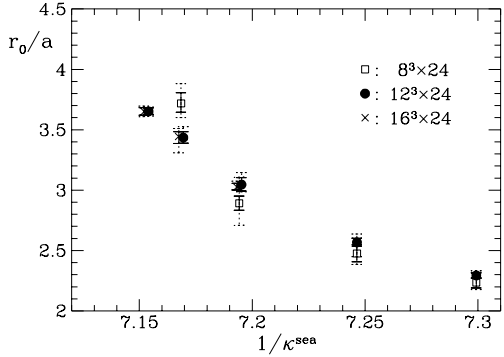,height=8.5cm,angle=0}
\vspace{-2.5truecm}
\caption{$r_0/a$ as a function of $1/\ksea$.  The solid errors
are statistical and the dashed errors are systematic, as explained 
in \protect\cite{UKQCD_C176}. \protect \label{fig:latspac}}
\end{center}
\vspace{-1.0truecm}
\end{figure}

To set the lattice spacing, we have used the scale $r_0$ determined from
\cite{Sommer}
\begin{equation}
F(r_0/a)(r_0/a)^2=1.65,\qquad r_0=0.49\ \mbox{fm},
\end{equation}
where $F(r)$ is the force between two static quarks, derived from the 
potential $V(r)$, which is in turn extracted in a standard way from
Wilson loops \cite{UKQCD_C176}.
In Full QCD $r_0$ is better suited than the string tension, since the string
is expected to break, and is also more reliable than the mass of the $\rho$,
since it can decay.
In fig.~\ref{fig:latspac} we show $r_0/a$ as function of $1/\ksea$ for the
three volumes.  Even though part of this strong dependence of $a$ on 
$\kappa_{\sea}$ could be reabsorbed by improving the coupling 
\cite{Martinelli}, it indicates the necessity to work at a fixed scale, 
whenever possible \cite{UKQCD_C202}. 
Moreover, while comparing $8^3\cdot 24$ and $12^3\cdot 24$ there is
a sizable difference at the lightest quark mass, the difference is
completely negligible between $12^3\cdot 24$ and $16^3\cdot 24$.  
We can summarise these results in a bound of $L/r_0\gtapprox 3.2$ 
above which finite-size effects in the static quark potential are
largely absent.  Given the approximate linear behaviour of $r_0/a$
at the three lightest quark masses, we have attempted a chiral extrapolation
for the two largest volumes.  In the chiral limit, we obtain
$a=0.121(2)$ fm and $a=0.122(2)$ fm, for $12^3\cdot 24$ and $16^3\cdot 24$.

\begin{figure}[t]
\hspace{-2.0truecm}
\vspace{-3.0truecm}
\begin{center}
\epsfig{figure=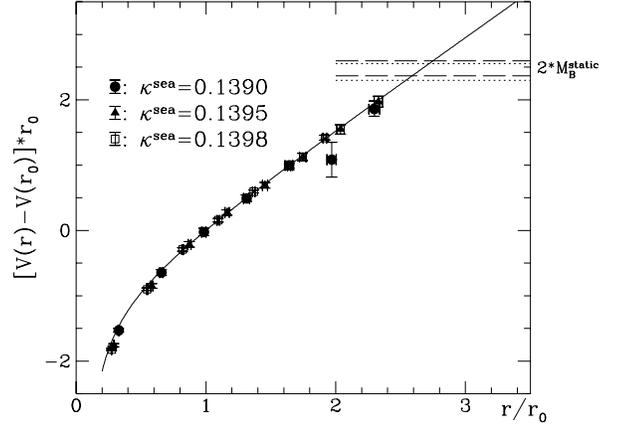,height=9cm,angle=0}
\vspace{-2.9truecm}
\caption{Rescaled potential at $V=12^3\cdot 24$ for the three lightest sea 
quark masses.\protect \label{fig:pot_scaled}}
\end{center}
\vspace{-1.0truecm}
\end{figure}

One can rescale the linear Coulomb ansatz of the potential
\begin{equation}
V(r)=V_0 + \sigma r - \frac{e}{r},\qquad e=\pi/12,
\end{equation}
to eliminate $\sigma$ as follows
\begin{equation}
[V(r)-V_0]r_0=(1.65-e)(r/r_0-1)-e(r_0/r-1).
\label{eq:(V-V0)r0}
\end{equation}
In fig.~\ref{fig:pot_scaled}, we show the rescaled potential 
for $12^3\cdot 24$ and for different values of $\ksea$.  The solid
line doesn't represent a fit but simply a plot of eq.~(\ref{eq:(V-V0)r0}).
In Full QCD, due to string breaking one expects the data to flatten out at 
large distances, below a value equal to twice the static meson mass.
These limiting values have been calculated for $\ksea=0.1390$ (dashes) and 
$\ksea=0.1395$ (dots).  With our present data, there is
not conclusive evidence of string breaking up to $r\simeq 2.5r_0$.
A more chiral sea quark mass or a better way of extracting the potential
are needed \cite{UKQCD_C176}.
At small distances, the data seems well described by the Coulombian part
even if the data for the lightest quark masses tend to lie somewhat below
the curve.  This seems to favour a value of $e>\pi/12$, as also observed
in \cite{SESAM} for unimproved Wilson fermions.  This is consistent
with the influence of the dynamical quarks on the potential through the
running of the strong coupling.  We interpret this as a sign of unquenching
and plan to carry out a quantitative analysis of the effect in a future
publication, using the ``correct'' value of $\csw$ \cite{UKQCD_C202}.

\section{Light Spectrum}

\begin{table}[h]
\vspace{-0.5truecm}
\begin{tabular}{cccccc}
\hline
$\kappa/$Volume & $8^3\cdot 24$ & $12^3\cdot 24$ & $16^3\cdot 24$ \\ 
\hline \hline
0.1370  & 0.857\err{6}{7}  & 0.855\err{2}{3}  &  \\
0.1380  & 0.829\err{9}{7}  & 0.825\err{4}{5}  &  \\
0.1390  & 0.769\err{10}{11}& 0.785\err{4}{7}  & 0.785\err{6}{7} \\
0.1395  & 0.710\err{16}{18}& 0.710\err{10}{10}& 0.719\err{7}{10} \\ 
0.1398  &                  & 0.674\err{9}{23} & 0.670\err{10}{13} \\ \hline
\end{tabular}
\caption{$m_{\rm PS}/m_{\rm V}$ ratios for the different volumes.
\protect\label{tab:masses}}
\vspace{-1.0truecm}
\end{table}

We have carried out a first complete analysis of the light hadron spectrum
on the entire dataset, to gather complementary information to that obtained
from the static quark potential.  
Amplitudes and masses have been obtained in a stardard way by correlated
least-$\chi^2$ fits to correlation functions.  We have used throughout
a double exponential fitting function, taking into account the backward
propagating state for mesons.  We have found that the influence of the
first excited state is not negligible as we approach the most chiral point.
In order to estimate the excitation, we have fitted simulaneously 
the correlator ``fuzzed'' both at sink and source, denoted FF, and
the correlator with local sink and source, denoted LL.   
The FF correlator allows a much faster isolation of the 
fundamental state than the LL correlator.

\begin{figure}[t]
\vspace{-2.5truecm}
\begin{center}
\epsfig{figure=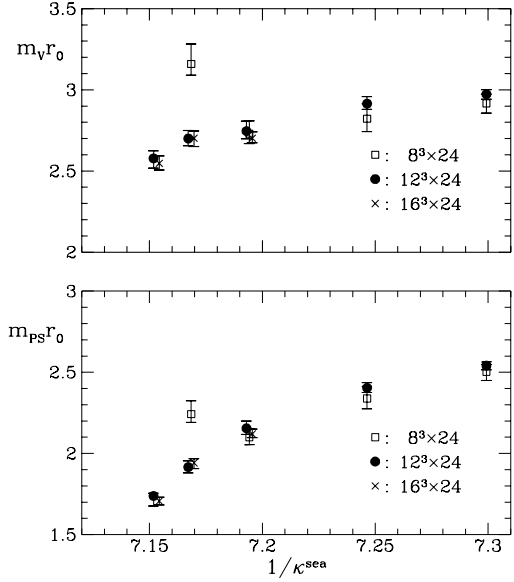,height=8.5cm,angle=0}
\vspace{-1.0truecm}
\caption{Meson masses for different volumes, in units of $r_0$.
\protect\label{fig:mesonsr0}}
\vspace{-1.0truecm}
\end{center}
\end{figure}

In fig.~\ref{fig:mesonsr0} we show the vector and pseudoscalar mesons,
as a function of $1/\ksea$, calculated at $\kval=\ksea$ 
for the different volumes.  To allow a significant comparison, we report
all masses in units of $r_0$, which reabsorbs the dependence of the lattice
spacing on $\ksea$.  The plots show pronounced finite-size effects
between $8^3\cdot 24$ and $12^3\cdot 24$ as we move towards the most
chiral point.  On the other hand, between $12^3\cdot 24$ and $16^3\cdot 24$
we find no significant discrepancy within statistical accuracy at all
values of the quark mass.  A qualitatively similar picture is bourne out of the
baryon sector.  This behaviour confirms the one found for the static
quark potential and we can extend to the mass spectrum the
bound for finite-size effects of $Lr_0\gtapprox 3.2$ already found there.

\begin{figure}[t]
\vspace{-3.5truecm}
\begin{center}
\epsfig{figure=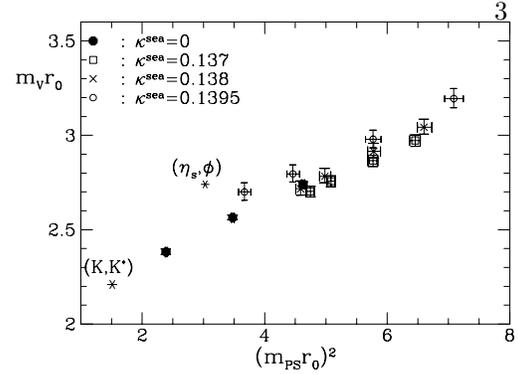,height=8.5cm,angle=0}
\vspace{-2.5truecm}
\caption{Vector vs pseudoscalar squared, in units of $r_0$, for 
$V=12^3\cdot 24$ at three values of $\ksea$ and comparison with quenched
($\ksea=0$).\protect\label{fig:V-PS2}}
\vspace{-1.0truecm}
\end{center}
\end{figure}

One way to investigate sea quark effects in the spectrum is through the
parameter $J$ \cite{UKQCD_J}, which only requires the input of a physical
V/PS mass ratio.  In Full QCD, we need to fix $\ksea$ and vary the valence
quark mass, interpreted as having strange flavour in a sea of light quarks.
However, our most chiral sea quark mass is only in the strange region.
Moreover, the errors are amplified by the fitting process to extract the
$dm_{\rm V}/dm_{\rm PS}^2$ slope, especially at the lightest mass.
In view of all this, it is not surprising that the values of $J$ show
no appreciable trend towards the experimental value as the sea quark mass
decreases, as shown in ref.~\cite{UKQCD_C176}.  Another way to look for 
dynamical effects, equivalent in spirit but which bypasses any fitting 
procedure, is to look at the $(m_{\rm V},m_{\rm PS}^2)$ plane directly for a 
shift in the data, calculated at fixed $\ksea$ and different $\kval$, 
as we approach the chiral limit.
In fig.~\ref{fig:V-PS2} we show such a plot and also report the quenched
result for comparison ($\ksea=0$), calculated at $\beta=5.7$ and 
$V=12^3\cdot 24$, which has a comparable lattice spacing. 
As the sea quark mass decreases, we do observe
a significant, albeit small, trend towards the point $(m_{\eta_s}^2,m_{\phi})$,
i.e.~the mesons whose flavour content resembles most closely that used in 
our simulation.  Moreover, at the lightest sea quark mass, we can assert
a significant shift compared to the quenched result.  To quantify 
this quenching effect, we do need to keep the scale constant,
an issue which will be addressed with the fully $O(a)$ improved action
\cite{UKQCD_C202}.

\end{document}